\documentclass{mn}[3]
\input psfig.sty
\begin{document}

\title[SdB+MS binaries from 1st RLOF]{The orbital periods of subdwarf B binaries produced by the first stable Roche overflow channel}
\author[X. Chen, Z. Han, J. Deca and Ph. Podsiadlowski]
{Xuefei Chen $^{\rm 1,2}$\thanks{E-mail: cxf@ynao.ac.cn} Zhanwen Han $^{\rm 1, 2}$  Jan Deca $^{\rm 3}$ and Philipp Podsiadlowski $^{\rm 4}$\\
$^1$ Yunnan Observatory, Chinese Academy of Sciences, Kunming, 650011, China\\
$^2$ Key Laboratory for the Structure and Evolution of Celestial Objects, Chinese Academy of Sciences, Kunming, 650011, China \\
$^3$ Centrum voor Plasma-Astrofysica, Katholieke University Leuven, Celestijnenlaan 200B, B-3001 Leuven, Belgium\\
$^4$ University of Oxford, Department of Astrophysics, Keble Road,
Oxford OX1 3RH}

\maketitle

\begin{abstract}\label{abstract}\label{firstpage}
Long-orbital-period subdwarf B (sdB) stars with main-sequence
companions are believed to be the product of stable Roche Lobe
overflow (RLOF), a scenario challenged by recent observations.  Here
we represent the results of a systematic study of the orbital-period
distribution of sdB binaries in this channel using detailed binary
evolution calculations. We show that the observed orbital-period
distribution of long-period sdB binaries can be well explained by this
scenario. Furthermore, we find that, if the progenitors of the sdB
stars have initial masses below the helium flash mass, the sdB
binaries produced from stable RLOF follow a unique mass -- orbital
period relation for a given metallicity $Z$; increasing the orbital
period from $\sim 400$ to $\sim 1100$\,d corresponds to increasing the
mass of the sdB star from $\sim 0.40$ to $\sim 0.49\,M_\odot$ for
$Z=0.02$. We suggest that the longest sdB binaries (with orbital
period $> 1100$\,d) could be the result of atmospheric RLOF. The mass --
orbital period relation can be tested observationally if the mass of
the sdB star can be determined precisely, e.g.\ from
asteroseismology. Using this relation, we revise the orbital period
distribution of sdB binaries produced by the first stable RLOF channel
for the best fitting model of Han et al (2003), and show that the
orbital period has a peak around 830\,d.

\end{abstract}


\section{INTRODUCTION}\label{1. INTRODUCTION}
In the Hertzsprung-Russell diagram, subdwarf B stars (sdBs) are
located between the main sequence (MS) and the white dwarf (WD)
cooling track at the blueward extension of the horizontal branch.
These objects play an important role in the study of stellar
evolution theory, asteroseismology, as distance indicators and for
Galactic structure and evolution (see the review of Heber 2009).
They are also thought to be sources of far-ultraviolet radiation
in early-type galaxies (Ferguson et al. 1991; Brown et al. 2000;
Han et al 2007).

SdBs are generally considered to be helium-core-burning stars with
extremely thin hydrogen envelopes ($<0.02\,M_\odot$) (Heber 1986;
Saffer et al. 1994). Han et al. (2002, 2003) developed a detailed
binary model for the formation of sdBs, which successfully explains
field sdBs and possibly {\it extreme horizontal-branch} stars in
globular clusters (Han 2008). They defined three types of formation
channels to produce sdBs: the first and second stable Roche Lobe
overflow (RLOF) channel for sdB binaries with long orbital periods,
the first and second common envelope (CE) ejection channel for binary
systems with short orbital periods, and the helium WD merger channel
for single sdBs. Han et al.\ (2003) showed that the contribution of
the second RLOF scenario is not significant. This means that most
long-orbital period sdB binaries formed via stable RLOF channels must
have MS companions. Such sdBs are hard to detect because of their
bright companions and slow variability with time. The first stable
RLOF channel therefore has remained untested due to the observational
absence of long-period sdB binaries with known orbital periods.

Recently, {\O}stensen \& Van Winckel (2012) published a first
sample of long-period sdB systems using data from the High
Efficiency and Resolution Mercator Echelle Spectrograph mounted at
the Mercator telescope in La Palma, Spain (Mercator sample
hereinafter). Deca et al. (2012) reported a 760 d period for PG
1018-047. In addition, Barlow et al. (2012) determined
precise orbital solutions for three systems by combining 6 years
of Hobby-Eberly Telescope data with recent observations from the
Mercator telescope. They present an up-to-date period histogram
for all known hot subdwarf binaries, suggesting a long period peak
to be around $500-1000$\,d. All these observations seem to
challenge the predictions of Han's conventional model, which
presented a period peak around 100 d and showed hardly any systems
beyond 500\,d.

This conflict results from a simplified treatment of the stable RLOF
channel in the binary population synthesis study in Han et al.(2003)
(hereinafter HPMM03), who were mainly interested in the CE channel,
since only short-period sdB binaries had been found at the
time. HPMM03 set the final masses to be the core masses of the giants
at the onset of RLOF and obtained the final orbital periods by
assuming that the lost mass takes away the angular momentum from the
companion star. Such a treatment is obviously too simple and can
result in significant discrepancies with observations; e.g.\ table 4
in Han et al. (2002) shows that sdBs are significantly heavier than
the core mass of their giant at the onset of RLOF, and that systems
can have orbital periods far beyond 500\,d if the giant is less than
$1.6\,M_\odot$. This illustrates the necessity for improving the
determination of the orbital period distribution for sdB binaries
evolved through the first RLOF channel, using full binary evolution
calculation rather than the simplified approach adopted in this
earlier study.

In fact, there are two subchannels for the first stable RLOF channel. (i)
If the primary has an initial zero-age main sequence mass (ZAMS)
below the helium flash mass (about $2\,M_\odot$ for Pop I), RLOF has
to occur near the tip of the first giant branch (FGB), which leads
to relatively long-period sdB+MS binaries ($>400$\,d as seen in
Table 4 of Han 2002). (ii) If the ZAMS mass of the primary is
larger than the helium flash mass and the system experiences
stable RLOF in the Hertzsprung gap (HG) or on the FGB, this also
leads to the formation of an sdB star, but with an orbital period
likely less than 100\,d as shown by Han et al. (2002), Chen \& Han
(2002, 2003). In this work we will mainly focus on the first
subchannel, as it is the more important one and the only one that is
relevant to explain the newly discovered systems.

In section 2, we introduce the stellar evolution code used and the
basic inputs needed for the study. The evolutionary consequences,
including the mass -- orbital period relation are presented in
section~3. In section~4 we discuss the sdB binaries with the longest
orbital periods ($>1100$\,d). The orbital period distribution is
presented in section~5, and our main conclusions are summarized in
section~6.

\section{Binary evolution calculations}
We use the stellar evolution code originally developed by Eggleton
(1971, 1972, 1973), updated with the latest input physics (Han et
al. 1994; Pols et al. 1995, 1998). We set the mixing length to
a local pressure scale height ratio $\alpha = 2.0$ ($\alpha =
l/H_{\rm p}$), the convective overshooting parameter, $\delta_{\rm
OV}$, is set to 0.12 (Schr\"{o}der et al. 1997), which roughly
corresponds to an overshooting length of 0.25$\,H_{\rm P}$. We
study four metallicities, i.e. $Z= 0.02$, 0.01, 0.004 and 0.001. The
opacity tables for these metallicities were compiled by Chen \&
Tout (2007) from Iglesias \& Rogers (1996) and Alexander \&
Ferguson (1994).

For each $Z$, the initial hydrogen mass fraction is obtained by $X
= 0.76-3Z$ (Pols et al. 1998). The donor stars can have four
initial masses, i.e. $M_{\rm 1i}=$ 0.8, 1.00, 1.26 or $1.6\,M_\odot$,
always lower than the helium flash mass for the respective
$Z$ value. The corresponding companion masses, $M_{\rm 2i}$, are
0.7, 0.9, 1.1 and $1.5\,M_\odot$, respectively, chosen to be less
than the donor star mass and at the same time ensuring stable
RLOF, i.e. the mass ratio $q$ ($\equiv$ the donor/the accretor) is
below the critical mass ratio $q_{\rm c}$. The initial orbital
period distribution is produced as follows: for orbital periods
$P_{\rm i}>100$\,d, we use intervals of 50\,d, whereas for shorter
periods a 20\,d interval is chosen, in order to ensure that sdB
stars can be produced with the corresponding assumptions. RLOF is
included in the model via the following boundary condition:
\begin{equation}
\frac{{\rm d}m}{{\rm d}t}=C\cdot {\rm Max}[0, (\frac{R}{R_{\rm
L}}-1)^3],
\end{equation}
where ${\rm d}m/{\rm d}t$ is the mass-loss rate of the giant due to
RLOF. The constant $C$ is set to $500\,M_\odot\, {\rm yr}^{-1}$ so
that RLOF can proceed steadily and the lobe-filling star overfills its
Roche lobe as necessary but never overfills its lobe by much.  Due to
convergence problems, we artificially fix the maximum value of ${\rm
  d}m/{\rm d}t$ at $0.5\times 10^{-4}\,M_\odot {\rm yr}^{-1}$ to make
sure that most of the model runs evolve to the helium flash (or helium
WD) successfully\footnote{The choice of this value has no influence on
  the final conclusions, in particular the sdB mass -- period
  relation, which is independent of both the mass-loss rate and the
  angular momentum loss, as shown in this work. }.

We define $\beta$ as the fraction of mass lost by the donor via
RLOF accreted onto the secondary. Three values of $\beta$  (0, 0.5
and 1) have been studied\footnote{Since the donor is now a giant,
RLOF is unlikely to be conservative. The value of $\beta=1$ is
included here for completeness of the calculations only.}. The
remaining part, $1-\beta$, is lost from the binary and takes away
the specific angular momentum (marked with a parameter $A$) as
pertains to the donor ($A=1$), or to the companion ($A=2$). A
Reimers-type wind with an efficiency of $\eta=1/4$ (Renzini 1981;
Iben \& Renzini 1983; Carraro et al. 1996) is included before the
onset of RLOF. The wind is lost from the binary by taking away the
specific angular momentum as pertains to the giant.

\section {Evolutionary consequences}
For each metallicity, the model produces three different outcomes:
sdB+MS binaries, He WD binaries (where the helium flash does not
occur) and red clump star binaries (RCS, where the helium flash
occurs when the donor has an envelope mass larger than
  0.0200$\,M_\odot$)\footnote{The maximum hydrogen envelope mass of
  sdB stars is related to the core mass and metallicity and is around
  $0.02\,M_\odot$ (Heber 2009). }. Figure 1 summarizes the remnant
mass ($M_{\rm f}$) versus the orbital period ($P$) for all of these
products. We use filled circles for sdB stars produced under the
various assumptions, with red, green, blue, light blue and purple
symbols respectively for ($\beta=0,A=2$); ($\beta=0.5,A=2$);
($\beta=0,A=1$), ($\beta=0.5,A=1$) and ($\beta=1$ (independent on $A$
for this case)). See also Table 1. Note that there is a clear $M_{\rm f}-P$
correlation for sdB and He WD binaries for a given metallicity, while
for the red clump binaries there is no clear trend, as one would expect
(see sect.~3.1 for further discussion).

\begin{figure*}
\centerline{\psfig{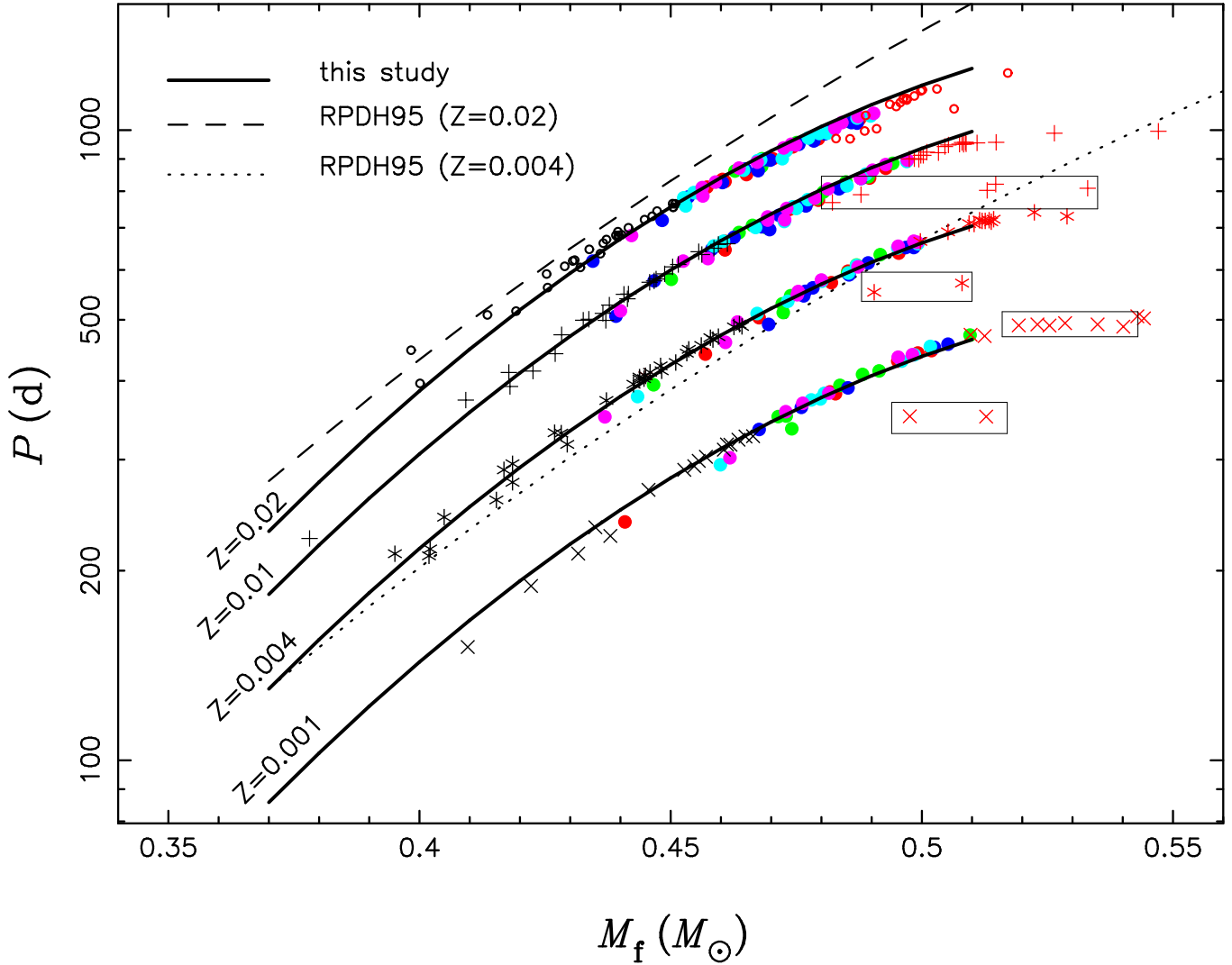}}
\caption{The remnant mass ($M_{\rm f}$) -- orbital period plane
for binaries produced in this study. The filled circles are sdB+MS
stars, where different colors represent those produced under
different assumptions. Red, green, blue, light blue and purple
filled circles represent $\beta=0,A=2$; $\beta=0.5,A=2$;
$\beta=0,A=1$, $\beta=0.5,A=1$ and $\beta=1$ (independent of $A$
for this case), respectively (see also Table 1). From top to
bottom, the metallicity is 0.02, 0.02, 0.004 and 0.001,
respectively. Open circles, pluses, asterisks and crosses are for
He WDs (low-mass end) and RCSs (high-mass end) at various
metallicities, $Z=0.02,0.01,0.004$ and 0.001, respectively.
Symbols enclosed in a rectangle are RCSs with similar orbital
periods but different sdB masses. The dashed and dotted curves
show the results from Rappaport et al. (1995)(RPDH95), i.e. Eq(6)
of that paper with $R_0=5500$ and $3300\,R_\odot$ for $Z=0.02$ and
0.004, respectively, while the solid curves are from the fitting
formulae of this study.
 \label{Fig1}}
\end{figure*}

\subsection{The $M_{\rm f}-P$ relation}
Giant donor stars generally have degenerate cores and follow the
well-known $M_{\rm c}-R$ (core mass -- radius) relation (Refsdal \&
Weigert 1971; Webbink, Rappaport \& Savonije 1983; Joss, Rappaport \&
Lewis, 1987). For a given core mass $M_{\rm c}$, we can obtain the
stellar radius $R$ of the giant and hence its Roche lobe radius,
$R_{\rm L}$, which is $\approx R$. This directly constrains the
orbital period and leads to a well-defined $M_{\rm c}-P$ relation
during stable RLOF. Roche lobe overflow stops when the donor star
envelope collapses, i.e.\ when $M_{\rm f} \simeq M_{\rm c}$.  This then
defines a $M_{\rm f}-P$ relation, which is independent of the angular
momentum loss during the mass-transfer phase. Since the stellar radius
decreases with decreasing metallicity, the orbital period $P$ will
decrease with decreasing $Z$ at a given $M_{\rm f}$ as shown
in Figure 1.

The $M_{\rm f}-P$ relation indicates that the remnant mass of the
giant and the final orbital period after RLOF are coupled
together. The way angular momentum is lost determines the remnant
mass; e.g., rapid angular momentum loss leads to a shorter RLOF phase
and a lower remnant mass for a given donor mass. This relation was
first proposed by Rappaport et al. (1995)(RPDH95 hereinafter) and
verified by many authors i.e. Tauris \& Savonije (1999), Nelson,
Dubeau \& MacCannell (2004), De Vito \& Benvenuto (2010, 2012),
although there are some differences in the detailed proposed forms of
this relation. These differences probably arise from differences
between the stellar evolution models (and codes) used.

RPDH95 obtained the $M_{\rm wd}-P$ relation (where $M_{\rm wd}$ is
the mass of the WD remnants after stable RLOF) in wide binary
radio pulsars, according to the $M_{\rm c}-R$ relation of single
giants. For comparison, we show the results of RPDH95 in Fig.~1.
Our result differs slightly from that of RPDH95 because of three
factors: (i) We used updated opacities, which directly affect the
stellar radius. Also, the Pop I models used in RPDH95 have
$2<\alpha<3$ and various He abundances, both influencing the
stellar radius and hence the final orbital period. (ii) The giant
expands when losing mass, resulting in a small amount of
divergence of the $M_{\rm c}-R$ relation derived from single
stars. (iii) The remnant mass of the giant is not exactly equal to
its core mass at the termination of RLOF. In general, a more
massive core has larger surface gravity, and the envelope starts
to collapse when it is still quite thick. Therefore the orbital
period increases more slowly (with $M_{\rm f}$ increasing) than
expected from the $M_{\rm c}-R$ relation for both Pop I and Pop II
stars.

For RCSs, the envelope is too thick ($>0.02\,M_\odot$) at the He flash
for the assumption of $M_{\rm f}\approx M_{\rm c}$ to be valid. In
this case, the orbital period is in fact determined by the core mass
of the RCS, as is also partly reflected in Fig.~1.  For example, there
are five pluses with orbital periods around 800\,d (enclosed in a
rectangle), but with masses in a wide range
($0.48-0.53\,M_\odot$). The ZAMS masses of the donors in the five
systems are $1.6\,M_\odot$ and their core masses are around
$0.45\,M_\odot$. There are several similar cases shown in this figure,
i.e.\ two lower asterisks, two lower crosses, crosses with mass larger
than $0.51\,M_\odot$ (enclosed in the rectangles of Fig.~1) etc. The
systems in a particular rectangle have similar orbital periods since
the donors have similar core masses at the He flash.

\subsection{The fitting formulae}
For convenience, we have fitted the $M_{\rm f}-P$ relation with a
polynomial formula for $Z=0.004$\footnote{We choose $Z=0.004$ because the
sample distribution in the mass range of $0.39-0.51\,M_\odot$ are
more uniform for this metallicity than others.}:
\begin{equation}
\log P=c_1+c_2M_{\rm f}+c_3M_{\rm f}^2,
\end{equation}
with $P$ in days and $M_{\rm f}$ in solar mass. The coefficients are
$c_1=-3.5605$, $c_2=22.6555$ and $c_3=-19.7851$, with a maximum error
of 1.1 percent in $P$. The orbital period for other metallicities at a
given $M_{\rm f}$ can be obtained by a simple vertical shift of Eq(2),
i.e. changing the value of the constant $c_1$. For $Z=0.02$, 0.01 and
0.001, we find $c_1=-3.3105$, $-3.4105$ and $-3.7505$, respectively,
with the maximum fitting error being 2.9, 1.1 and 1.4 percent in $P$,
respectively\footnote{The overall best fit for $Z=0.02$ is
  $c_1=-1.8203$, $c_2=16.5762$, and $c_3=-13.6420$ in Eq(2), with a
  maximum error of 1.2 percent in $P$.}. This means that for different
metallicities, the orbital period increases at a similar rate (with
$M_{\rm f}$), which is to be expected since all the binaries have
evolved by stable RLOF. The fitting formulae are plotted in Fig.~1 as
solid curves.

\begin{table*}
\small \caption{The sdB mass ($M_{\rm sdB}$) and orbital period ($P$)
  of sdB+MS binaries in our studies. The first two columns show the
  ZAMS masses of the donors ($M_{\rm 1i}$) and the initial orbital
  periods ($P_{\rm i}$), respectively. The companion masses are 0.7,
  0.9, 1.1 and $1.5\,M_\odot$ for $M_{\rm 1i}=0.79, 1.00, 1.26$ and
  $1.6\,M_\odot$, respectively. The initial orbital periods are
  different for different cases to ensure that sdB stars are produced
  with the corresponding assumptions. The interval of $P_{\rm i}$ is
  $\sim 50$\,d when $P_{\rm i}>100$\,d and 20\,d when $P_{\rm
    i}<100$\,d. $\beta$ is the fraction of the mass lost from the
  donor via RLOF accreted onto the secondary, and $A$ is a label for
  the angular momentum loss; it is 1, if the mass lost from the system
  takes away the same specific angular momentum as the donor, and 2,
  if it the specific angular momentum of the accretor.  The results
  are independent of $A$ for the case $\beta=1.0$ (conservative mass
  transfer). The masses and orbital period are in units of solar masses
  and days, respectively.} \tiny
\begin{tabular}{rcccrcccrcccrcccccc}
\hline $Z=$&0.02&&&$Z=$&0.01&&&$Z=$&0.004&&&$Z=$&0.001&\\
 \hline
$M_{\rm 1i}$& $P_{\rm i}$&$M_{\rm sdB}$&$P$&$M_{\rm 1i}$&
$P_{\rm i}$&$M_{\rm sdB}$&$P$&$M_{\rm 1i}$& $P_{\rm i}$&$M_{\rm sdB}$&$P$&$M_{\rm 1i}$& $P_{\rm i}$&$M_{\rm sdB}$&$P$\\
 \hline
$\beta=0$&$A$=2&\\
 0.79 & 400.0 & 0.4608 &  829.2 &  0.79 & 350.0 & 0.4683 &  704.3 &  0.79 & 250.0 & 0.4676 &  504.0 &  0.79 & 200.0 & 0.4828 &  381.6\\
 0.79 & 450.0 & 0.4703 &  900.6 &  0.79 & 400.0 & 0.4794 &  774.1 & 0.79 & 300.0 & 0.4820 &  572.6 &  0.79 & 250.0 & 0.5019 &  446.8\\
 0.79 & 500.0 & 0.4794 &  968.5 &  0.79 & 450.0 & 0.4896 &  839.3 &  0.79 & 350.0 & 0.4954 &  637.9 & 1.00 & 150.0 & 0.4952 &  430.0\\
 1.00 & 250.0 & 0.4603 &  835.1 &  1.00 & 200.0 & 0.4608 &  665.9 & 1.00 & 200.0 & 0.4853 &  597.3 & 1.26 &  80.0 & 0.4817 &  384.2\\
 1.00 & 300.0 & 0.4743 &  940.9 &  1.00 & 250.0 & 0.4775 &  772.0 & 1.60 &  40.0 & 0.4569 &  441.1 &  1.26 & 100.0 & 0.4992 &  442.9\\
 1.00 & 351.0 & 0.4876 & 1041.0 &  1.00 & 300.0 & 0.4928 &  870.1 &      &       &        &        & 1.58 &  20.0 & 0.4409 &  238.9\\
 1.26 & 150.0 & 0.4572 &  812.0 &  1.26 & 150.0 & 0.4736 &  748.3 & \\
 1.26 & 200.0 & 0.4786 &  975.2 &  1.60 &  60.0 & 0.4608 &  645.7 & \\
 1.60 &  80.0 & 0.4651 &  850.1 & \\
 &&&&\\
$\beta=0.5$&$A$=2&\\
 0.79 & 450.0 & 0.4601 &  831.7 & 0.79 & 400.0 & 0.4692 &  715.6 & 0.79 & 300.0 & 0.4722 &  530.2 & 0.79 & 200.0 & 0.4730 &  351.9\\
 0.79 & 500.0 & 0.4689 &  897.0 & 0.79 & 450.0 & 0.4792 &  779.4 & 0.79 & 350.0 & 0.4852 &  593.1 & 0.79 & 250.0 & 0.4915 &  414.9\\
 1.00 & 300.0 & 0.4628 &  860.9 & 0.79 & 500.0 & 0.4893 &  843.3 & 0.79 & 400.0 & 0.4978 &  654.5 & 0.79 & 300.0 & 0.5096 &  473.0\\
 1.00 & 351.0 & 0.4751 &  955.2 & 1.00 & 250.0 & 0.4663 &  706.0 & 1.00 & 200.0 & 0.4739 &  546.0 & 1.00 & 150.0 & 0.4837 &  393.6\\
 1.00 & 400.0 & 0.4868 & 1044.0 & 1.00 & 300.0 & 0.4804 &  797.6 & 1.00 & 250.0 & 0.4917 &  634.3 & 1.26 &  80.0 & 0.4714 &  351.1\\
 1.26 & 200.0 & 0.4682 &  900.7 & 1.00 & 350.0 & 0.4942 &  885.5 & 1.26 & 150.0 & 0.4870 &  610.9 & 1.26 & 100.0 & 0.4882 &  409.6\\
 1.26 & 250.0 & 0.4853 & 1033.0 & 1.26 & 150.0 & 0.4636 &  688.0 & 1.60 &  40.0 & 0.4466 &  394.4 & 1.58 &  40.0 & 0.4741 &  335.8\\
 1.60 &  80.0 & 0.4529 &  761.0 & 1.26 & 201.0 & 0.4849 &  827.5 & 1.60 &  60.0 & 0.4724 &  513.6 &      &       &        &\\
 1.60 & 100.0 & 0.4681 &  875.7 & 1.60 &  60.0 & 0.4501 &  579.9 &      &       &        &        &      &       &        &\\
      &       &        &        & 1.60 &  80.0 & 0.4691 &  699.5 &      &       &        &        &      &       &        &\\
 &&&&&\\
$\beta=0$&$A$=1&\\
 0.79 & 500.0 & 0.4603 &  825.9 & 0.79 & 450.0 & 0.4681 &  702.3 & 0.79 & 350.0 & 0.4765 &  545.7 &  0.79 & 250.0 & 0.4853 &  389.9\\
 0.79 & 551.0 & 0.4698 &  896.3 &  0.79 & 500.0 & 0.4769 &  757.9 &  0.79 & 400.0 & 0.4855 &  589.2 & 0.79 & 300.0 & 0.5052 &  457.4\\
 0.79 & 600.0 & 0.4785 &  961.3 & 1.00 & 350.0 & 0.4626 &  675.9 &  0.79 & 450.0 & 0.4984 &  651.3 & 1.00 & 200.0 & 0.4760 &  363.2\\
 0.79 & 650.0 & 0.4871 & 1026.0 &  1.00 & 400.0 & 0.4716 &  732.5 & 1.00 & 300.0 & 0.4783 &  561.9 &  1.00 & 250.0 & 0.5025 &  452.1\\
 1.00 & 400.0 & 0.4535 &  783.5 &  1.00 & 450.0 & 0.4879 &  838.1 &  1.00 & 350.0 & 0.4893 &  615.6 & 1.26 & 150.0 & 0.4676 &  334.9\\
 1.00 & 450.0 & 0.4648 &  867.5 &  1.26 & 300.0 & 0.4582 &  647.3 & 1.26 & 250.0 & 0.4768 &  554.7 & \\
 1.00 & 500.0 & 0.4755 &  949.4 & 1.26 & 350.0 & 0.4765 &  764.7 &  1.26 & 301.0 & 0.4968 &  651.2 & \\
 1.00 & 550.0 & 0.4860 & 1028.0 & 1.26 & 400.0 & 0.4846 &  816.2 &  1.60 & 200.0 & 0.4695 &  491.6 & \\
 1.26 & 400.0 & 0.4689 &  897.8 &  1.26 & 450.0 & 0.4835 &  806.9 & \\
 1.26 & 451.0 & 0.4817 &  996.2 &  1.60 & 200.0 & 0.4391 &  506.7 & \\
 1.26 & 500.0 & 0.4807 &  986.7 &  1.60 & 250.0 & 0.4561 &  611.9 & \\
 1.60 & 250.0 & 0.4345 &  619.7 &  1.60 & 300.0 & 0.4697 &  695.2 & \\
 1.60 & 300.0 & 0.4483 &  719.0 &       &       &        &        & \\
 &&&&&\\
$\beta=0.5$&$A$=1&\\
 1.60 & 350.0 & 0.4674 &  862.4 &  0.79 & 450.0 & 0.4669 &  700.7 & 0.79 & 350.0 & 0.4733 &  535.0 &  0.79 & 250.0 & 0.4798 &  374.5\\
 0.79 & 500.0 & 0.4553 &  795.6 & 0.79 & 500.0 & 0.4750 &  751.0 &  0.79 & 400.0 & 0.4854 &  593.4 & 0.79 & 300.0 & 0.4960 &  429.9\\
 0.79 & 551.0 & 0.4645 &  863.7 &  0.79 & 550.0 & 0.4851 &  815.9 &  0.79 & 450.0 & 0.4992 &  660.4 &  1.00 & 200.0 & 0.4806 &  382.3\\
 0.79 & 600.0 & 0.4730 &  927.6 & 0.79 & 600.0 & 0.4970 &  891.4 &  1.00 & 300.0 & 0.4801 &  575.6 &  1.00 & 250.0 & 0.5017 &  453.8\\
 0.79 & 650.0 & 0.4812 &  989.1 & 1.00 & 350.0 & 0.4605 &  666.3 &  1.00 & 350.0 & 0.4952 &  650.2 & 1.26 & 150.0 & 0.4780 &  373.1\\
 0.79 & 700.0 & 0.4898 & 1053.0 &  1.00 & 400.0 & 0.4773 &  775.9 &  1.26 & 200.0 & 0.4672 &  511.8 & 1.26 & 200.0 & 0.4985 &  440.0\\
 1.00 & 400.0 & 0.4526 &  781.0 & 1.00 & 450.0 & 0.4889 &  851.4 &  1.26 & 251.0 & 0.4870 &  609.8 &  1.58 &  80.0 & 0.4599 &  294.5\\
 1.00 & 450.0 & 0.4673 &  894.3 & 1.26 & 250.0 & 0.4586 &  655.0 &  1.60 & 100.0 & 0.4434 &  377.7 & \\
 1.00 & 500.0 & 0.4774 &  971.6 &  1.26 & 300.0 & 0.4743 &  755.8 & \\
 1.00 & 550.0 & 0.4871 & 1045.0 &  1.26 & 350.0 & 0.4888 &  850.8 & \\
 1.26 & 300.0 & 0.4546 &  796.0 & 1.60 & 200.0 & 0.4726 &  716.2  &\\
 1.26 & 350.0 & 0.4679 &  897.1 & \\
 1.26 & 400.0 & 0.4796 &  987.6 & \\
 1.60 & 200.0 & 0.4530 &  758.1 & \\
 1.60 & 251.0 & 0.4722 &  900.7 & \\
 &&&&\\
$\beta=1.0$&&\\
 0.79 & 500.0 & 0.4589 &  827.1 &  0.79 & 450.0 & 0.4693 &  720.0 &  0.79 & 350.0 & 0.4752 &  547.9 & 0.79 & 250.0 & 0.4815 &  382.6\\
 0.79 & 551.0 & 0.4672 &  889.1 &  0.79 & 500.0 & 0.4787 &  780.0 &  0.79 & 400.0 & 0.4873 &  606.7 &  0.79 & 300.0 & 0.4981 &  440.3\\
 0.79 & 600.0 & 0.4749 &  947.7 &  0.79 & 550.0 & 0.4878 &  838.7 &  0.79 & 450.0 & 0.4991 &  663.8 &  1.00 & 150.0 & 0.4729 &  357.5\\
 0.79 & 650.0 & 0.4827 & 1006.0 &  0.79 & 600.0 & 0.4971 &  896.1 &  1.00 & 200.0 & 0.4633 &  495.9 &  1.00 & 200.0 & 0.4953 &  436.3\\
 0.79 & 700.0 & 0.4904 & 1063.0 &  1.00 & 301.0 & 0.4693 &  728.1 &  1.00 & 250.0 & 0.4800 &  578.4 & 1.26 & 100.0 & 0.4763 &  368.7\\
 1.00 & 351.0 & 0.4636 &  870.2 &  1.00 & 350.0 & 0.4813 &  806.4 &  1.00 & 300.0 & 0.4953 &  654.2 &  1.58 &  40.0 & 0.4618 &  302.0\\
 1.00 & 400.0 & 0.4740 &  949.4 &  1.00 & 400.0 & 0.4928 &  881.4 &  1.26 & 150.0 & 0.4754 &  554.4 & \\
 1.00 & 451.0 & 0.4841 & 1028.0 &  1.26 & 150.0 & 0.4525 &  619.6 &  1.26 & 200.0 & 0.4984 &  666.5 & \\
 1.26 & 200.0 & 0.4563 &  810.8 & 1.26 & 201.0 & 0.4731 &  751.2 &  1.60 &  40.0 & 0.4369 &  350.8 & \\
 1.26 & 250.0 & 0.4725 &  936.9 & 1.26 & 250.0 & 0.4902 &  863.8 &  1.60 &  60.0 & 0.4609 &  460.4 & \\
 1.26 & 300.0 & 0.4873 & 1051.0 &  1.60 &  60.0 & 0.4400 &  516.7 & \\
 1.60 &  80.0 & 0.4422 &  680.5 &  1.60 &  80.0 & 0.4574 &  625.3 & \\
 1.60 & 100.0 & 0.4564 &  786.1 & 1.60 & 100.0 & 0.4727 &  720.8 & \\
 \hline \label{sdb}

\end{tabular}
\end{table*}

\subsection{PG\,1018--047}
PG 1018-047 is a long-orbital-period sdB binary whose
MS companion and orbital parameters have been studied in detail, allowing
direct constraints of the stable RLOF channel. It has a mid-K
(K3--K5) MS companion with an orbital period of $759\pm5.8$\,d and a
mass ratio (MS/sdB) of $1.6\pm0.2$. If PG\,1018--047 has indeed a
circular orbit, it is a good candidate to have formed via first
stable Roche lobe overflow (Deca et al. 2012). However, for this
system to have evolved within a Hubble time with close-to-conservative
mass transfer, the present K star must have been no
heavier than $\sim 0.3-0.4\,M_{\odot}$, leading to a contradiction
with the initial mass ratio of the binary. We refer to Deca et al.
(2012) for an elaborate discussion.

Previous studies have shown that non-conservative RLOF (Han et al.
2002; Chen \& Han 2008) or a small mass ratio (the ratio of donor to
accretor mass) or both most likely lead to stable RLOF, indicating
that PG 1018-047 is likely the product of non-conservative RLOF, with
$Z=0.01-0.02$ (from Fig.~1). This metallicity indicates PG 1018-047
possible to be a thin disk star. On the other hand, its location,
R.A(J2000): 10h 21m 10.50s, DEC(J2000): $-04^\circ 56^{'} 19.3^{''}$
(Deca 2010) seems to suggest a halo star. A high-resolution
spectroscopic study of PG\,1018-047, including metallicity
constraints, is well under way and should help to address this issue
in more detail in the future.

\section{The longest period ($>1100$\,d) sdB binaries}
Figure 1 shows that sdB binaries produced by the first stable
Roche lobe overflow channel have a narrow mass range and hence a
limited range of possible orbital periods. For $Z=0.02$, the
orbital period changes from $\sim 600$ to 1100\,d as the sdB mass
increases from $\sim 0.43$ to $\sim 0.49\,M_\odot$\footnote{The
lower limit of orbital period can be as low as $\sim 400$\,d when
the donor's ZAMS mass is $1.9\,M_\odot$, see Table 4 of Han et al.
(2002).}. The longest system in our calculations has $P=1063$\,d
with a mass of $0.4904\,M_\odot$ for this metallicity (see Table 1).
According to the $M_{\rm f}-P$ relation, the upper limit of the
orbital period is determined by the maximum sdB mass, which can be
estimated by the sum of the maximum core mass of the giant on the
FGB ($M_{\rm c}^{\rm tip}=0.4746\,M_\odot$, see Table 1 in Han et
al. 2002) and the maximum envelope mass allowed for sdB stars
($0.02\,M_\odot$). The maximum sdB mass is then $0.4946\,M_\odot$,
which corresponds to an orbital period of 1135 d.

There are three sdB+MS binaries with orbital periods longer than
1100 d, i.e.\ BD+$29^{\circ}3070$, BD-$7^{\circ}5977$ and
PG1317+123. Their orbital periods are 1160, 1194 and 1179 d,
respectively({\O}stensen \& Van Winckel 2012, Barlow et al. 2012).
If we consider observational errors for the three objects (67, 79
and 12 d, respectively) and the uncertainties on stellar radius in
the model calculations, the three systems are consistent with the
$M_{\rm f}-P$ relation.

In our binary evolution, the RLOF switches on (off) sharply when
the radius of the donor is larger (smaller) than its Roche lobe
radius. However this assumption does not take into account the
possibility of material (atmosphere) outside the photospheric
stellar radius. This means that, when the stellar radius
approaches the Roche Lobe radius, some atmospheric material will
already overflow the Roche lobe and be transferred to the
companion (Ritter 1988, Pastetter \& Ritter, 1989; Podsiadlowski
et al.\ 2002, Chen et al. 2010). This is referred to as
atmospheric RLOF. In this case, the Roche Lobe is larger than the
stellar radius during mass transfer, resulting in a larger
separation and thus a longer orbital period. If the underfilling
factor is 10 percent as suggested by S-type symbiotics with
ellipsoidal variability (Rutkowski, Mikolajewska \& Whitelock
2007; Otulakowska-Hypka, Mikolajewska \& Whitelock 2013), the
Roche lobe radius $R_{\rm L}$, and hence the separation $a$, are
1.1 times larger than before. The orbital period $P$ will then be
1.17 times longer because $P \propto a^{3/2}$. Some sdB binaries
with orbital periods of $940-1025$\,d will then be shifted into
the $1100-1200$\,d range, significantly increasing the number of
sdB+MS binaries with orbital period longer than 1100 d in the
model\footnote{The effect of atmospheric RLOF discussed here is
just an informed estimate, and  the corresponding period
distribution shown in Fig.~2 is re-scaled from that obtained
without atmospheric-RLOF. Detailed calculations with
atmospheric-RLOF implemented may reveal some differences.}.

Furthermore, there may be other channels, e.g.\ a tidally enhanced stellar wind
channel or an envelope ejection channel at the tip of the FGB (Han et al.
2010) that can also lead to sdB+MS binaries with very long orbital
periods.

\section{The orbital period distribution}
In order to investigate the orbital period distribution of sdB
binaries from the stable RLOF, we have performed a Monte Carlo
simulation similar to those of HPMM03 but with the latest version
of the binary population synthesis code. The simulation is for a
Population I population and the required parameters in the Monte
Carlo simulation are the same as those in HPMM03 i.e.\ the star
formation rate is taken to be constant over the last 15\,Gyr, the
initial mass function of the primary are from a simple
approximation of Miller \& Scalo (1979), the initial mass-ratio
($q$, the ratio of the secondary to primary mass) distribution is
also set to be constant ($n(q)=1,0<q\le 1$), all stars are assumed
to be members of binaries and the distribution of separations($a$)
is constant in $\log a$ for wide binaries and falls off smoothly
at close separations. The critical mass ratio for dynamically
stable mass transfer between giants and their companions, $q_{\rm
crit}$, the common-envelope ejection efficiency, $\alpha_{\rm
CE}$, and the thermal contribution to the CE ejection,
$\alpha_{\rm th}$, are set to be 1.5, 0.75 and  0.75,
respectively, the same as that of the best fitting model of HPMM03
(set 2 in that paper).

As described in section 1, there are two subchannels, one for
progenitors with ZAMS mass less than $2\,M_\odot$, one for
progenitors with ZAMS mass greater than $2\,M_\odot$, that lead to
the formation of sdB long-period binaries. For the subchannel of
progenitors with ZAMS mass less than $2\,M_\odot$, the sdB
binaries follow the unique mass -- orbital period relation. The
sdB mass produced from stable RLOF is determined by the details of
the mass transfer process. For simplicity, we assume that the
masses of the sdB stars produced in this way are uniformly
distributed between $M_{\rm sdB}^{\rm min}$ and $M_{\rm sdB}^{\rm
max}$ if the ZAMS mass of the donor is less than $2\,M_\odot$ (the
helium flash mass for Population I), where $M_{\rm sdB}^{\rm
min}=0.4570, 0.4552, 0.4550, 0.4425$ and $0.4064\,M_\odot$,
$M_{\rm sdB}^{\rm max}=M_{\rm c}^{\rm
  tip}+0.02=0.4946,0.4927,0.4923, 0.4801$ and $0.4287\,M_\odot$, for
the donor's mass of 0.8, 1.0, 1.26, 1.6 and $1.9\,M_\odot$,
respectively (see Tables 1 and 4 in Han et al. 2002). For other
donor masses less than $2\,M_\odot$, $M_{\rm sdB}^{\rm min}$ and
$M_{\rm sdB}^{\rm max}$ are linearly interpolated from these five
stars. The orbital period is then obtained from the fitting
  formulae in sect.~3.2.

For the subchannel of progenitors with ZAMS mass greater than
$2\,M_\odot$, the sdB mass and orbital period do not follow the
sdB mass -- orbital period relation, but the orbital period
depends on both the sdB mass and the amount of angular momentum
loss, which are obtained in a way similar to that of HPMM03. The
sdB mass is taken from the detailed binary evolution models of Han
et al.\ (2000) when RLOF begins in the HG or set to be the core
mass of the donor at the onset of RLOF when RLOF starts on the FGB
(since there no detailed study of this channel yet, using full
binary evolution calculations), and 50 percent of the mass lost
from the donor is lost from the system, taking away the same
specific angular momentum as pertains to the system. The orbital
periods of sdB stars in this case are very sensitive to the
assumptions about angular momentum loss, which will be discussed
at the end of this section.

\begin{figure}
\centerline{\psfig{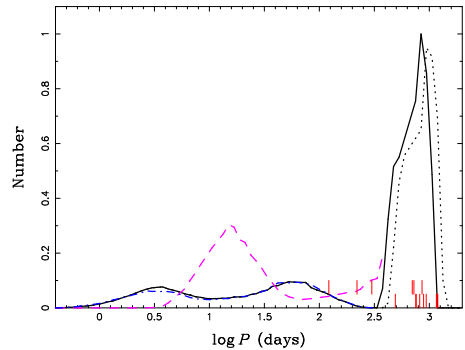}}
\caption{The distribution of orbital periods of sdB stars from the
  first stable RLOF. Different curves are for different assumptions
  for systems with donor ZAMS masses larger than $2\,M_\odot$; solid:
  the final mass of the donor is obtained from Han et al. (2000) and
  the mass lost from the system takes away the specific angular
  momentum as pertains to the system [the standard set]; dash-dotted:
  the final mass of the donor is obtained from the non-conservative
  binary calculations of Chen \& Han (2002), in comparison to the
  standard set; dashed: the mass lost takes away the specific angular
  momentum as pertains to the donor in comparison to the standard set
  (the dashed and dash-dotted curves are plotted only for $\log P{\rm
    (d)}<2.6$ for clarity). The dotted curve is for atmospheric-RLOF
  i.e.\ the orbital periods of sdB stars from the He flash models are
  multiplied by 1.17 (see sect.\ 4 for details). Short ticks along and
  above the X-axis indicate the positions of sdB+MS stars in the
  observational sample for solved and unsolved long-orbital sdB stars,
  respectively ({\O}stensen \& Van Winckel 2012, Barlow et al. 2012,
  Deca et al. 2012).
 \label{Fig2}}
\end{figure}

Fig.~2 shows the orbital period distribution of sdB binaries from the
first stable RLOF channel. The distribution (the solid curve) has two
parts, separated by a well-defined gap. The left-hand part contains
systems where the ZAMS mass of the donor is larger than $2\,M_\odot$
and the right-hand part are systems with ZAMS donor masses less than
$2\,M_\odot$. The gap is caused by the sharp drop of the radius at the
tip of FGB from stars with ZAMS masses somewhat smaller than
$2\,M_\odot$ (the helium flash mass for Pop I) relative to stars with
ZAMS masses somewhat greater than $2\,M_\odot$ (see Han et
al. 2002). The left-hand part has two peaks: the first peak ($\sim
4$\,d) is from systems undergoing RLOF during the HG while the second
peak ($\sim 63$\,d) comes from systems undergoing RLOF on the FGB. The
second peak was originally buried in the main peak of the dashed curve
of Fig.~10 in HPMM03, and appears when sdB stars from systems with
ZAMS donor mass smaller than $2\,M_\odot$ have been shifted to longer
orbital periods. The donors of binaries that produce the second peak
have non-degenerate cores on the FGB, and their behaviour is somewhat
different from those with degenerate cores, e.g.\ there is no sudden
collapse of the envelope at the end of RLOF. Thus, sdB stars from
these systems will not exactly follow the mass-orbital period relation
obtained from the donors with degenerate cores. Detailed binary
evolution calculations are necessary to improve modeling of such
systems, which we leave for a future study.

The right-most peak in the period distribution lies between $\sim
400-1100$\,d with a peak around 830 d. Most observed long-orbital
period sdB stars fall into this range, i.e.\ can be well explained
by this channel. Taking into account the possibility of
atmospheric RLOF may extend this range to orbital periods as long
as $\sim 1600$\,d (the dotted curve), and could easily explain the
longest period sdB binaries ($>1100$\,d) known to-date.  There are
three sdB stars with orbital periods less than 400 d but longer
than 100 d. They are unsolved systems and further observations are
necessary to confirm their orbital periods. They may not come from
the first stable RLOF for Pop I stars if their orbital periods are
in this range. The first CE ejection channel is a possible channel
can produce sdB+MS binaries with orbital periods between 100 and
400\,d. HPMM03 pointed out that the upper limit of the orbital
period distribution from the first CE channel could be as long as
400\,d in the extreme case of $\alpha_{\rm CE}=\alpha_{\rm th}=1$.

As shown in sect.~3, the orbital period of sdB stars from the
first stable RLOF with ZAMS donor mass smaller than $2\,M_\odot$
is only dependent on the sdB mass. The whole right-most peak of
the period distribution in Fig.~2 then cannot be affected by the
assumed angular momentum loss. But the left-hand part of the
distribution is significantly influenced by this assumption. If we
assume that the mass lost takes away the same specific angular
momentum as pertains to the donor (significantly smaller than that
of system) rather than to the system, the orbital periods of sdBs
become longer, and the first orbital period peak moves to $\sim
16$\,d while the second peak now overlaps with the right-hand part
(the dashed curve). In general, as one would expect, a smaller
amount of angular momentum loss leads to longer orbital periods.
So, the observed orbital period distribution for sdB stars with
masses significantly diverging from $\sim 0.48\,M_\odot$ (sdB
stars with mass around 0.48$\,M_\odot$ are likely from the first
CE channel) can be used to constrain the angular momentum loss
during RLOF in the HG directly. Since the donors do not suffer a
sudden collapse of the envelope at the end of RLOF if their ZAMS
masses are larger than $2\,M_\odot$, the orbital period
distribution of sdB star mass from these systems is not sensitive
to the sdB mass. We see from Fig.~2 that, if we adopt the final
masses of the donors obtained from non-conservative calculations
(e.g.\ Chen \& Han, 2002), the orbital period distribution (the
dash-dotted curve) is similar to that obtained from conservative
calculations.

\section{Summary}
In this paper we have investigated the orbital periods of sdB stars
produced from the first stable RLOF with ZAMS donor star masses
somewhat lower than the helium flash mass. Using detailed binary
stellar evolution calculations, we have determined a unique mass --
orbital period relation for sdB stars and WD binaries produced in this
way for various metallicities. This relation is a direct consequence
of the core mass -- radius relation of giant stars and the sudden
collapse of the giants at the end of RLOF. Binaries with red clump
stars (RCSs) do not follow this relation because their envelopes are
too thick at the He flash and the donors have not yet suffered a
sudden collapse of their envelopes. The final orbital period of RCSs
is then determined by the core mass of the donor at the He flash, not
the mass of the RCSs. The mass -- orbital period relation can be
verified from observations if the sdB mass could be precisely
determined, e.g.\ from asteroseismology.

Implementing this mass -- orbital period relation into binary
population synthesis code, we re-evaluated the distribution of orbital
periods of sdB stars from the first stable RLOF for a Population I
distribution. There is a wide orbital period range i.e.\ from several
days to $\sim 1100$\,d for sdB stars produced from the first stable
RLOF. If the ZAMS mass of the donors are less than $2\,M_\odot$, the
orbital period increases from $\sim 400$ to $\sim 1100$\,d as the sdB
mass increases from $\sim 0.40$ to $\sim 0.49\,M_\odot$. The period
peak is around 830 d, corresponding to a sdB mass of $\sim
0.46\,M_\odot$. Most observed long-orbital-period sdB binaries are
located in this range and are therefore well explained by this
formation channel. The longest sdB binaries (with orbital period $>
1100$\,d) are likely a consequence of atmospheric RLOF, while the sdB
stars with orbital periods in the range of $100-400$\,d may come from
the first CE ejection channel.

\section*{Acknowledgments}
This work is partly supported by the NSFC (Nos. 10973036,
11173055, 11033008 and 11003003), the CAS (No. KJCX2-YW-T24 and
the Talent Project of Western Light )and the Talent Project of
Young Researchers of Yunnan province (2012HB037).


\begin{thebibliography}{}
\bibitem[Alexander \& Ferguson (1994)]{af94}
Alexander D.R., Ferguson J.W., 1994, ApJ, 437, 879
\bibitem[Barlow et al. (2012)]{bar12}
Barlow et al., 2012, ApJ, 758, 58
\bibitem[Brown et al. (2000)]{brown00}
Brown T. M., Bowers C. W., Kimble R. A., Sweigant A. V., Ferguson
H. C., 2000, ApJ, 532, 308
\bibitem[Carraro et al. (1996)]{car96}
Carraro G., Girardi L., Bressan A., Chiosi C., 1996, A\&A, 305,
849
\bibitem[Chen \& Han (2002)]{chen02}
 Chen X., Han Z., 2002, MNRAS, 335, 948
\bibitem[Chen \& Han (2003)]{chen03}
 Chen X., Han Z., 2003, MNRAS, 341, 662
\bibitem[Chen \& Han (2008)]{chen08}
 Chen X., Han Z., 2008, MNRAS, 387, 1416
\bibitem[Chen \& Tout (2007)]{chen07}
 Chen X., Tout C. A., 2007, ChJAA, 7(No.2), 245
\bibitem[Chen et al. (2010)]{chen10}
 Chen X., Podsiadlowski Ph., Mikolajewska J., Han Z., 2010,
 in Kolagera V. and van der Sluys M., eds, AIP Conf. Proc. 314,
 International Conference on Binaries, p.59
 \bibitem[Deca 2010)]{deca10}
 Deca J., 2010, PG1018-047: The longest period subdwarf B binary
 (Unpubkished master's thesis), Katholieke Universitteit Leuven,
 Leuven, Belgium
 \bibitem[Deca et al. (2012)]{deca12}
 Deca J. et al., 2012, MNRAS, 421, 2798
\bibitem[De Vito \& Benvenuto (2010)]{vito10}
 De Vito M. A., Benvenuto O. G., 2010, MNRAS, 401, 2552
\bibitem[De Vito \& Benvenuto (2012)]{vito12}
 De Vito M. A., Benvenuto O. G., 2012, MNRAS, 421, 2206
\bibitem[Eggleton (1971)]{egg71}
 Eggleton P. P., 1971, MNRAS, 151, 351
\bibitem[Eggleton (1972)]{egg72}
 Eggleton P. P., 1972, MNRAS, 156, 361
\bibitem[Eggleton (1973)]{egg73}
 Eggleton P. P., 1973, MNRAS, 163, 279
\bibitem[Ferguson et al. (1991)]{fer91}
 Ferguson H. C. et al., 1991, 382, L69
 \bibitem[Han (2008)]{han08}
 Han Z., 2008, A\&A, 484, L31
 \bibitem[Han et al. (1994)]{han94}
 Han Z., Podsiadlowski Ph., Eggleton P. P., 1994, MNRAS, 270,
 121
  \bibitem[Han, Podsiadlowski and Lynas-Gray (2007)]{han07}
 Han Z., Podsiadlowski Ph., Lynas-Gray A. E., 2007, MNRAS, 380,
 1098
 \bibitem[Han et al. (2002)]{han02}
 Han Z., Podsiadlowski Ph., Maxted P. F. L., Marsh T. R., Ivanova N., 2002, MNRAS, 336,
 449
\bibitem[Han et al. (2003)]{han03}
 Han Z., Podsiadlowski Ph., Maxted P. F. L., Marsh T. R., 2003, MNRAS,
 341, 669 (HPMM03)
 \bibitem[Han, Tout \& Eggleton (2000)]{han00}
 Han Z., Tout C. A., Eggleton P. P., 2000, MNRAS, 319, 215
\bibitem[Han, Chen \& Lei (2010)]{han10}
 Han Z., Chen X., Lei Z., 2010, AIPC, 1314, 85
 \bibitem[Heber (1986)]{heber86}
 Heber U., 1986, A\&A, 155, 33
\bibitem[Heber (2009)]{heber09}
 Heber U., 2009, ARA\&A, 47, 211
\bibitem[Iben \& Renzini (1983)]{ir83}
 Iben I. Jr, Renzini A., 1983, ARA\&A, 21, 271
 \bibitem[Iglesias \& Rogers (1996)]{ir96}
 Iglesias C.A., Rogers F.G., 1996, ApJ, 464, 943
\bibitem[Joss, Rappaport \& Lewis (1987)]{joss87}
 Joss P. C., Rappaport S., Lewis W., 1987, ApJ, 319, 180
\bibitem[Miller \& Scalo (1979)]{miller79}
 Miller G. E., Scalo J. M., 1979, ApJS, 41, 513
 \bibitem[Nelson, Dubeau \& MacCannell (1996)]{ndm96}
 Nelson L. A., Dubeau E., MacCannell K. A., 2004, ApJ, 616, 1124
 \bibitem[ {\O}stensen \& Van Winckel (2012) ]{roy12}
 {\O}stensen R., Van WinckelH., 2012, in David Kilkenny,
 C. Simon Jeffery, and Chris Koen eds, ASP Conf. Ser. Vol.452,
 Fifth Meeting on Hot Subdwarf Stars and Related Objects, p.163
  \bibitem[Otulakowska-Hypka, Mikolajewska \& Whitelock
(2013)]{otu13} Otulakowska-Hypka M., Mikolajewska J., Whitelock
P.A., 2013, P. A. Woudt and V. A. R. M. Ribeiro, eds, STELLA
NOVAE: FUTURE AND PAST DECADES, ASP Conference Series, in press.
\bibitem[Pastetter \& Ritter (1989)]{past89}
 Pastetter L., Ritter H., 1989, A\&A, 214, 186
\bibitem[Podsiadlowski, Rappaport \& Pfahl (2002)]{pod02}
 Podsiadlowski Ph., Rappaport S., Pfahl E. D., 2002, ApJ, 565,
 1107
\bibitem[Pols et al. (1998)]{pol98}
 Pols O. R., Schr\"{o}der K. P., Hurley J. R., Tout C. A.,
 Eggleton P. P., 1998, MNRAS, 298, 525
 \bibitem[Pols et al. (1995)]{pol95}
 Pols O. R., Tout C. A., Eggleton P. P., Han Z., 1995, MNRAS, 274,
 964
\bibitem[Rappaport et al. (1995)]{rap95}
 Rappaport S., Podsiadlowski Ph., Joss P. C., Di Stefano R., Han
 Z., 1995, MNRAS, 273, 731 (RPDH95)
 \bibitem[Refsdals \& Weigert (1971)]{ref71}
 Refsdal S., Weigert A., 1971, A\&A, 13, 367
 \bibitem[Renzini (1981)]{ren81}
 Renzini A., 1981, in Chiosi C., Stalio R., eds, Effects of Mass Loss on Stellar
Evolution. Reidel, Dordrecht, p. 319
 \bibitem[Ritter (1988)]{ritter88}
 Ritter H., 1988, A\&A, 202, 93
 \bibitem[Rutkowski, Mikolajewska \& Whitelock (2007)]{rut07}
Rutkowski A., Mikolajewska J., Whitelock P.A., 2007, Baltic Astr.
16, p. 49
 \bibitem[Saffer et al. (1994)]{saff94}
 Saffer R. A., Bergeron P., Koester D., Liebert J., 1994, ApJ,
 432, 351
\bibitem[Schr\"{o}der et al. (1997)]{sch97}
Schr\"{o}der K. P., Pols O. R., Eggleton P. P., 1997, MNRAS, 285,
696
\bibitem[Tauris \& Savonije(1999)]{tau99}
 Tauris T. M., Savonije G. J., 1999, A\&A, 350, 928
\bibitem[Webbink, Rappaport \& Savonije(1983)]{web83}
 Webbink R. F., Rappaport S., Savonije G. J., 1983, ApJ, 270, 678


\end{thebibliography}
\end{document}